# Monitoring the rotary motors of single $F_oF_1$-ATP synthase by synchronized multi channel TCSPC

N. Zarrabi[a], M. G. Düser[a], S. Ernst[a], R. Reuter[a], G. D. Glick[b], S. D. Dunn[c], J. Wrachtrup[a], M. Börsch*[a]

[a]3rd Institute of Physics, University of Stuttgart, Pfaffenwaldring 57, 70550 Stuttgart, Germany;
[b]Department of Chemistry, University of Michigan, Ann Arbor, MI, USA 48109-1001
[c]Department of Biochemistry, University of Western Ontario, London, Ontario, Canada N6A 5C1

**ABSTRACT**

Confocal time resolved single-molecule spectroscopy using pulsed laser excitation and synchronized multi channel time correlated single photon counting (TCSPC) provides detailed information about the conformational changes of a biological motor in real time. We studied the formation of adenosine triphosphate, ATP, from ADP and phosphate by $F_oF_1$-ATP synthase. The reaction is performed by a stepwise internal rotation of subunits of the lipid membrane-embedded enzyme. Using Förster-type fluorescence resonance energy transfer, FRET, we detected rotation of this biological motor by sequential changes of intramolecular distances within a single $F_oF_1$-ATP synthase. Prolonged observation times of single enzymes were achieved by functional immobilization to the glass surface. The stepwise rotary subunit movements were identified by Hidden Markov Models (HMM) which were trained with single-molecule FRET trajectories. To improve the accuracy of the HMM analysis we included the single-molecule fluorescence lifetime of the FRET donor and used alternating laser excitation to co-localize the FRET acceptor independently within a photon burst. The HMM analysis yielded the orientations and dwell times of rotary subunits during stepwise rotation. In addition, the action mode of bactericidal drugs, i.e. inhibitors of $F_oF_1$-ATP synthase like aurovertin, could be investigated by the time resolved single-molecule FRET approach.

## 1. INTRODUCTION

$F_oF_1$-ATP synthases catalyze the formation of adenosine triphosphate, ATP, from ADP and phosphate. $F_oF_1$-ATP synthases are embedded in the inner mitochondrial membrane, the thylakoid membrane of chloroplasts, and in the plasma membrane of bacteria. To synthesize ATP the enzyme from *Escherichia coli* utilizes the difference of the electrochemical potential of protons across a lipid membrane, i.e. a pH difference plus an electric potential generated by ion concentration differences[1]. Proton translocation through the membrane-bound $F_o$ part of ATP synthase drives a rotary motion of the ring of *c* subunits with respect to the non-rotating subunits *a* and *b*. This rotation is transmitted to the γ and ε subunits of the $F_1$ part (see Fig. 1) which accommodates three catalytic binding sites for ATP or ADP plus phosphate as well as three non-catalytic binding sites, respectively. The binding sites are located at the interface of the β and α subunits[2]. Rotation of the γ and ε subunits induces the conformational changes in the binding sites which allow for binding of the substrates, catalytic action and release of the products[3]. The bacterial $F_oF_1$-ATP synthases catalyze not only ATP synthesis but also the reverse chemical reaction of ATP hydrolysis. Thereby the γ and ε subunits are forced to rotate in the opposite direction in the $F_1$ part[4], and protons are pumped through the coupled $F_o$ part. According to the rotary catalysis mechanism small molecule inhibitors of ATP synthesis or hydrolysis are expected to block the rotary motions. Different mechanisms have to be considered like direct binding to the nucleotide or proton binding sites in a competitive way, binding to subunit domains which have move, or by weakening the non-covalent bounds between subunits and uncoupling catalysis from proton translocation.

*m.boersch@physik.uni-stuttgart.de; phone (49) 711 6856 4632; fax (49) 711 6856 5281; www.m-boersch.org

Subunit rotations of the two distinct motors of $F_oF_1$-ATP synthase have been investigated with biochemical[5] and spectroscopical[6] methods. Direct videomicroscopic evidence for ATP-driven rotation was provided using surface-immobilized single $F_1$ parts[4] followed by the unraveling of substeps using nanometer sized beads as a marker of rotation[7]. The rotational mode of the $F_o$ motor has been studied by confocal single-fluorophore anisotropy[8] and by videomicroscopy of single beads attached to $F_o$ [9, 10].

We apply a single-molecule Förster-type fluorescence resonance energy transfer (FRET) approach[11-16] using two fluorophores to monitor the rotary motions within a single $F_oF_1$-ATP synthase. Distance changes between the two fluorophores, one covalently attached to a non-rotating subunit and the second one bound to a rotating subunit, can be resolved in real time with millisecond time resolution and sub-nanometer precision. Rotation will result in a sequence of distance changes between these two fluorophores, and, accordingly, the FRET efficiency is expected to change stepwise. Previously determination of FRET efficiency levels and dwell times were done manually in the time trajectories of double-labeled ATP synthases, which were reconstituted into liposomes and were freely diffusing through the confocal single-molecule detection volume.

To overcome the limitations of a potentially biased FRET data analysis, a robust software-based approach has to be developed. Furthermore, artifacts caused by spectral fluctuations of the FRET fluorophores have to be identified and excluded from the analysis of conformational dynamics. This can be achieved by simultaneous measurements of FRET donor and acceptor with two pulsed lasers and synchronized multi channel TCSPC detection.

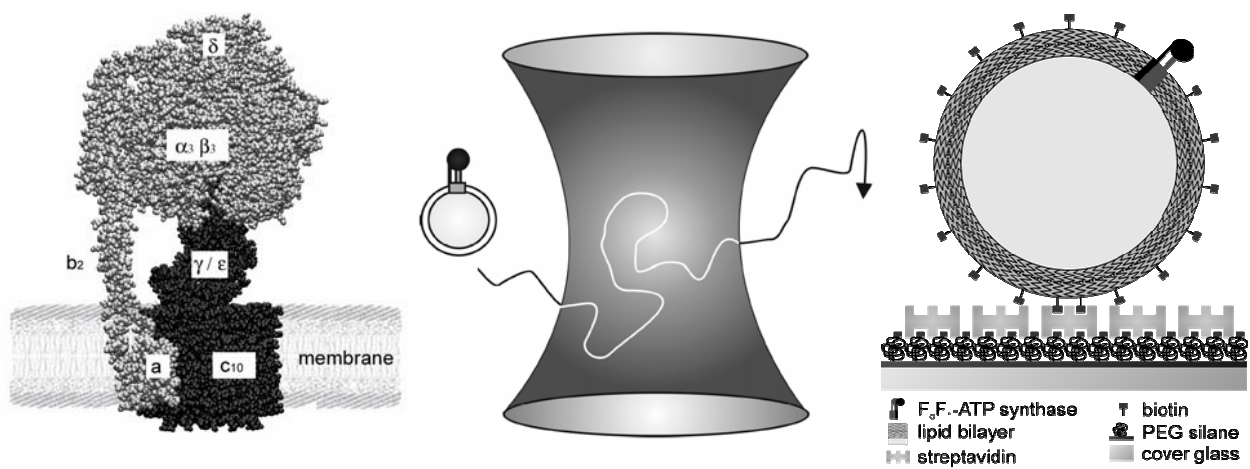

Fig. 1. Left, model of $F_oF_1$-ATP synthase embedded in the lipid membrane of a vesicle. The $F_1$ motor consists of the non-rotating subunits $\alpha_3\beta_3\delta$ (grey) and the rotating $\gamma$ and $\varepsilon$ subunits (black). The $F_o$ motor comprise the static subunits *a* and $b_2$ (grey) and the rotating ring of ten *c* subunits (black). For the time resolved FRET experiments in solution, the donor fluorophore EGFP was directly bound to subunit *a* and the acceptor dye Alexa-568 to the $\varepsilon$ subunit[15]. For the surface-immobilized FRET measurements the FRET donor sulforhodamine-B was bound to the top of one β subunit and the acceptor Cy5 to the γ subunit[12]. Middle, confocal FRET approach to monitor rotation in single liposome-bound $F_oF_1$-ATP synthases, which are freely diffusing in buffer solution through the laser focus. Right, surface-immobilized $F_oF_1$-ATP synthase in a liposome. The enzyme is attached by the streptavidin-bound liposome to a PEG passivated glass. FRET efficiency changes of the $F_oF_1$-ATP synthase are measured by confocal sample scanning.

## 2. EXPERIMENTAL PROCEDURES

### 2.1. FRET-labeled $F_oF_1$-ATP synthases in liposomes

We investigated two different FRET labeled $F_oF_1$-ATP synthases. The preparation procedure for one of the FRET-labeled $F_oF_1$-ATP synthases from *Escherichia coli* has been published[12]. Briefly, the $F_oF_1$-ATP synthases used for surface immobilization with biotin-containing lipid vesicles was selectively labeled with sulforhodamine B sulfonyl fluoride (Fluka) as the FRET donor at one lysine (lys-4) residue of only one of the non-rotating β subunits. The FRET acceptor Cy5 (GE Healthcare) was attached to a genetically introduced reactive cysteine (cys-106) at the rotating γ subunit. Both fluorophores are bound to the $F_1$ part. The double-labeled $F_1$ parts were re-assembled with $F_o$ parts which had been reconstituted to lipid vesicles with a diameter of about 100 nm. In contrast to our previous studies, the lipid bilayer contained about 1 % of a third lipid component (Biotin-DHPE, Invitrogen) for the binding of the liposomes to a streptavidin coated surface as shown in Fig. 1.

For the time resolved FRET measurements with alternating laser excitation we used a new genetic fusion of the 'enhanced Green Fluorescent Protein', EGFP, to the C terminus of subunit *a* in $F_o$, that has been constructed by Y. Bi (University of Western Ontario, Canada; to be published elsewhere). The FRET acceptor was bound to a cysteine at residue position 56 of the ε subunit[14,15]. $F_oF_1$-ATP synthase was purified from bacterial membranes as described[1,11,13]. Attaching Alexa-568 (Molecular Probes) to a single cysteine of detergent-solubilized $F_oF_1$ was accomplished a 4° C using the maleimide derivatives of the fluorophores. Rhodamine labeling efficiency was determined by UV-VIS absorption spectroscopy using the EGFP absorbance as an internal concentration reference. After labeling, ATP hydrolysis rates were measured as a control. Enzymes were reconstituted in an excess of preformed liposomes to assure the ratio of a single enzyme per liposome. The FRET-labeled $F_oF_1$-ATP synthases in liposomes were stored as 2 or 5 μl aliquots at -80° C.

### 2.2. Confocal microscope setup

Single-molecule FRET measurements were accomplished on a custom-designed confocal microscope based on an inverted Olympus IX 71. Fiber coupled ps-pulsed laser sources were available at 488 nm (PicoTa 490, up to 80 MHz repetition rate, Picoquant) and 638 nm (LDH-P-635, Picoquant) which could be triggered externally[15]. Continuous-wave laser excitation could be achieved by an argon ion laser (Spectra Physics), a tunable Ti:sapphire laser (Spectra Physics), or an frequency doubled Nd:YAG at 532 nm (Compass 315M, Coherent). However, for the recent FRET experiments we attached rhodamine dyes like Alexa-568 or sulforhodamine B to the $F_oF_1$-ATP synthases. In order to excite these fluorophores we used a continuous-wave DPSSL laser at 561 nm (Cobolt Jive, 25 mW, kindly provided by von Gegerfelt Photonics) in combination with a clean-up filter (HQ 560/60, AHF analysentechnik).

Laser beams at 488 nm and/or 561 nm were expanded for small, diffraction limited excitation volumes or compressed for the solution measurements, respectively, re-directed by a dual band dichrioc filter (HC dual line beam splitter 488/633-638, AHF) and focussed by a water immersion objective (UPlanSApo 60xW, 1.2 N.A., Olympus). Fluorescence of FRET donor and acceptor was detected by two avalanche photodiodes simultaneously (AQR-14, Perkin Elmer). For surface-immobilized FRET measurements, fluorescence was detected after passing a 50 μm pinhole in the donor channel from 575 nm to 640 nm (HQ 610/70, AHF) and in the acceptor channel above 640 nm (separated by a dicroic HQ640LP with additional blocking of scattered laser excitation by HQ595LP, AHF). For the solution measurements with dual excitation at 488 nm and 561 nm, fluorescence passed a 150 μm pinhole and was detected in the donor channel between 497 nm and 567 nm (HQ 532/70; AHF) and in the acceptor channel for λ > 595 nm separated by a dichroic (HQ575LP, AHF). Single photons were registered by two synchronized TCSPC cards (SPC-152, Becker & Hickl, kindly provided) and, simultaneously, by a multi channel counter card for the FRET imaging (NI-PCI 6602, National Instruments). Sample scanning was accomplished by a x-y piezo scanner plus a piezo objective positioner (Physik Instrumente), and the digital controller was addressed by custom-made software written in LabView. Images and FRET efficiencies in each pixel were analyzed with the software 'ODA-analyzer'.

## 2.3. Dual laser excitation for single-molecule fluorescence lifetime measurements with variable repetition rates

For FRET acceptor excitation at 561 nm in an alternating dual-laser excitation scheme the laser light was switched by an acousto-optic modulator, AOM, (model 3350-192, Crystal technologies). The AOM had rising and falling times of at least 10 ns, which led to relatively long laser pulses of about 24 ns. Due to the prolonged pulse duration the effective peak excitation power was much lower compared to a ps-pulsed laser source which reduced photobleaching of the fluorophores. A voltage controllable microwave oscillator (ZX95-400, 200 to 380 MHz, Mini-Circuits) was tuned to a frequency of 350 MHz according to the specifications of the AOM. The microwave was fed through a TTL controlled switch (ZASWA-2-50DR, Mini-Circuits). A microwave amplifier (ZHL-1-2W, 2 W, Mini-Circuits) with an upstream connected attenuator (ZX73-2500, Mini-Circuits) increased the microwave power.

The AOM, the PicoTa 490 and the SPC-152 TCSCP cards had to be triggered independently. The pulse sequences as shown in Fig. 2 were generated with an Arbitrary Waveform Generator (AWG2041, Tektronix, 1 GSa/s). The device provides an 8-bit digital equivalent on 8 ECL outputs of the analog arbitrary waveform with a time resolution of 1 ns. The digital ECL signals were transformed to TTL signals by a custom-made transformer in order to control the switch for the AOM, to trigger the pulsed laser and to synchronize the TCSCP cards.

Alternating the two lasers would drop the excitation rate of the FRET donor (83.3 MHz) to 33 % (27.7 MHz). The FRET information could be optimized if the photon count rates of the two fluorophores EGFP and Alexa-568 were maximized, that is, if the exciting laser for the FRET donor is more often pulsed during one period. We generated a series of six consecutive pulses for the donor exciting laser and one pulse for the acceptor probing. The 12 ns pulse to the AOM turned the acceptor probing laser on, the following 12 ns pause switched the 561 nm laser off. The 10 ns rising and falling times of the AOM led to the smooth shape of the first peak in the microtime histogram of the TCSPC cards. Subsequently six consecutive pulses to the PicoTA laser with an interval of 12 ns created the six fluorescence lifetime peaks in the microtime histogram. With this pulse sequence the maximum excitation rate of the FRET donor of 83.3 MHz is reduced to 75% (62.5 MHz) only.

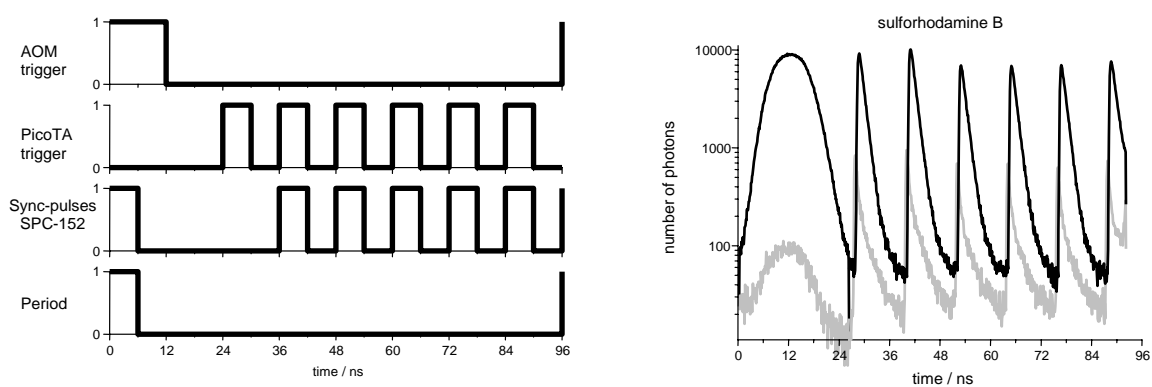

Fig. 2. TTL pulse sequences for multiple FRET donor excitation. Left, trigger pulse sequence for the AOM, the PicoTA 490 and the TCSPC cards synchronization. Right, microtime histogram of the photons as detected from the two APDs in one period (donor channel in grey, acceptor channel in black) using sulforhodamine B in $H_2O$.

The photons corresponding to the multiple 488 nm excitations in the microtime histogram contained the lifetime information of the FRET donor and had to be combined for fitting. Therefore six consecutive but time shifted synchronization pulses were sent to the TCSPC cards. The resulting microtime histogram is shown in Fig. 3.

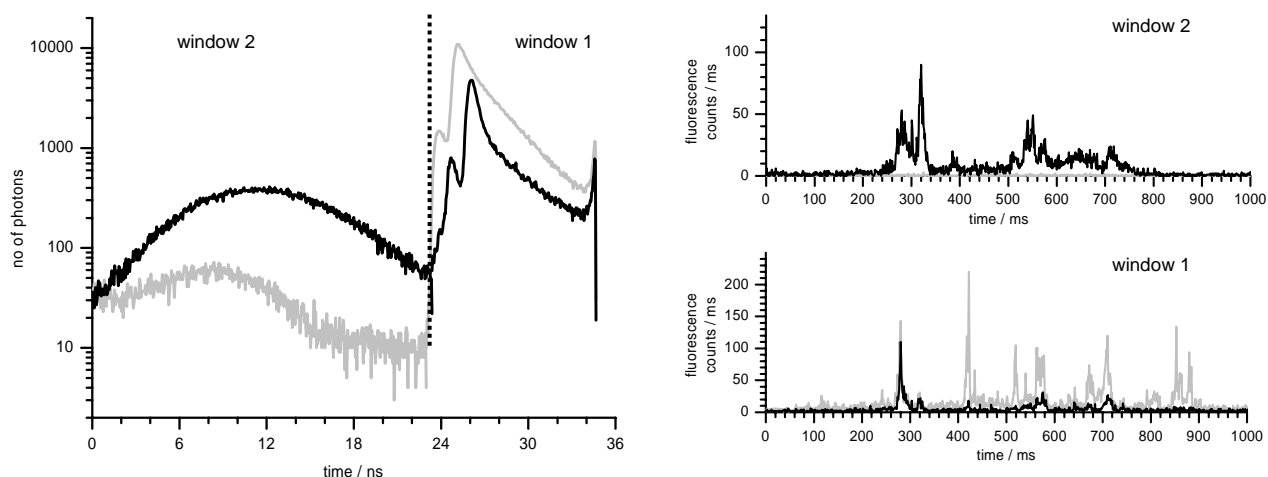

Fig. 3. Left, principle of the time windowed single photon counting measurement. By separating the signal of the two APDs according to their microtime information, four timetraces were generated. Time window 1 contained the fluorescence of donor (grey traces) and acceptor (black traces) by FRET donor excitation at 488 nm, and time window 2 the acceptor fluorescence by FRET acceptor excitation (black traces) at 561 nm. FRET data of freely diffusing $F_oF_1$-ATP synthases in liposomes are shown. Binning time is 1 ms.

Following the separation of the microtime histograms of both APDs into two parts, the first window 1 provided the FRET distance information, and the second window 2 contained the information about the presence or absence of the acceptor dye. Accordingly four time traces were generated: time trace 1 contained FRET donor photons (directly excited), time trace 2 provided the photons from the acceptor dye due to FRET, time trace 3 comprised fluorescence photons from the FRET donor dye excited by the red-shifted 561 nm laser (typically very few), and time trace 4 showed the fluorescence intensity of the directly excited acceptor dye.

## 3. RESULTS

### 3.1. FRET measurements of single $F_oF_1$-ATP synthases bound to the cover glass surface during ATP hydrolysis

It was anticipated that detailed information of the rotary catalysis mechanism could be gained from observation of immobilized $F_oF_1$-ATP synthases during ATP hydrolysis. Previous attempts had failed using non-specifically adsorbed bovine serum albumine, BSA, to cover the glass surface and to attach a streptavidin layer on top for binding of biotin lipid containing liposomes. The $F_oF_1$-ATP synthases in liposomes were found to be attached to the streptavidin layer and the existence of three different FRET efficiencies indicated the expected conformations of the enzymes. However, in the presence of ATP no changes in FRET efficiencies and, accordingly, no subunit rotation corresponding to catalytic activity of the enzymes could be detected.

Surface-immobilized FRET samples provide well-defined FRET levels from constant fluorescence intensities because the Brownian motion of particles through the laser focus is lacking. Those FRET trajectories are suitability for Hidden Markov models. Only with immobilized ATP synthases a slower turnover at low ATP concentration is possible to measure as well as the influence of inhibitors that might slow down rotation. Furthermore the rotary subunit motion in immobilized $F_oF_1$-ATP synthases can be compared directly with the rotation data of similarly immobilized $F_1$ parts. This is more general FRET approach. Additionally pulsed excitation can be used and high time resolution FRET data can be obtained by confocal sample scanning.

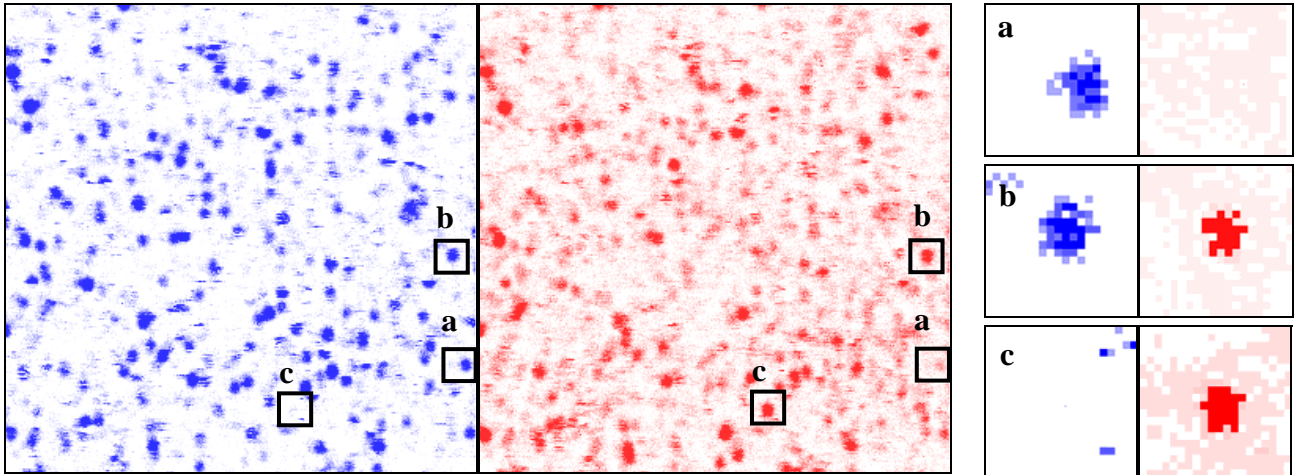

Fig. 4. Confocal FRET images of single immobilized $F_oF_1$-ATP synthases in the presence of ATP. Scanning directions were from bottom to top and from left to right, resolution is 300 x 300 pixel (20 x 20 μm), 4 ms pixel time. Left, FRET donor intensity, middle, FRET acceptor intensity. Right, three selected spots (20 x 20 pixel) showing three different conformations of the enzyme: (a) low FRET state, (b) medium FRET state, and (c) high FRET state.

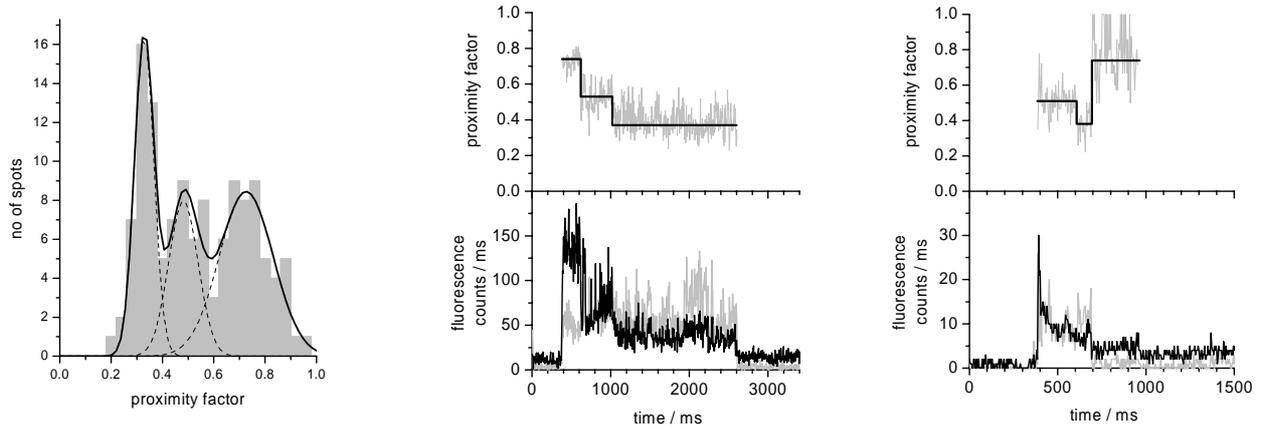

Fig. 5. Left, histogram of the proximity factors of identified $F_oF_1$-ATP synthases from Fig. 4. Middle and right, time trajectories of the FRET efficiency changes of two $F_oF_1$-ATP synthases in the presence of 1 mM ATP. The order of FRET efficiency changes shown in the proximity factor trace is high-FRET level → medium → low → high as measured in buffer solution[12].

Functional binding of $F_oF_1$-ATP synthases in liposomes as shown in Fig. 5 was achieved by carefully passivating the glass surface. As a first step the plasma purified cover glass was silanized to provide reactive amino groups for the subsequent chemistry. Modified PEG was bound covalently to the amino groups containing a small fraction of biotin but most of the PEG was carboxy-terminated. A streptavidin layer was added to the biotin and unbound streptavidin was removed by washing with buffer solution. The biotin-lipid containing liposomes with single $F_oF_1$-ATP synthases were allowed to bind to the modified cover glass and binding to the surface was measured simultaneously by repeated confocal sample scanning. As shown in Fig. 4, individual immobilized $F_oF_1$-ATP synthases appeared as bright spots in the FRET donor image and the FRET acceptor image.

As a measure of the FRET efficiency the proximity factor P was calculated

$$P = \frac{I_A}{I_A + I_D} \quad (1)$$

with $I_A$ and $I_D$, background corrected fluorescence intensities in the FRET acceptor or FRET donor channel, respectively. A background of 2 kHz in each channel was substracted.

The images in Fig. 4 provided 125 FRET spots which were summed up to the histogram in Fig. 5. The brightness per spot was the added fluorescence intensity in circular area with a diameter of eight pixels around the center of each marked spot. From those intensities, the proximity factor values were calculated for each spot. Only spots with a constant proximity factor value were taken into account. Three maxima in the histogram indicated the three stopping positions of the γ subunits of the $F_oF_1$-ATP synthase, with P = 0.33 (σ=0.08) for the low-FRET orientation (probably inculding the FRET donor only population), P = 0.48 (σ=0.11) for the medium-FRET orientation, and P = 0.73 (σ=0.21) for the high-FRET orientation.

Within some of the $F_oF_1$-ATP synthases spots in the FRET image the FRET efficiency changed during the scan indicating rotary motion of the γ subunit during ATP hydrolysis. To achieve the high time resolution for monitoring the FRET efficiency changes we used the synchronized TCSPC electronics for photon recording. At first a 20 x 20 μm FRET image of the immobilized $F_oF_1$-ATP synthases was measured with low excitation intensity (10 μW) and a fast 2 ms pixel time. The images were analyzed automatically to find the center of the spots, which were identified by a minimum intensity threshold in one detector channel. Up to 150 spots were found in one image. Then we increased the excitation power to 30 μW. The piezo scanner moved the sample to the laser focus and simultaneously fluorescence was recorded for 10 seconds for each spot. The FRET efficiency trajectories (binned to 1 ms) were calculated from the TCSPC data files and analyzed.

Fig. 5 shows two FRET trajectories of rotating $F_oF_1$-ATP synthases. The observation of one enzyme started with a high-FRET orientation (P = 0.74) of the γ subunit which was maintained for 236 ms. Afterwards, the γ subunit rotated within 1 ms to a medium-FRET orientation (P = 0.53) and remained in this conformation for 400 ms. A fast jump to the low-FRET orientation with P = 0.38 was accomplished and preserved for 1580 ms. The second enzyme started in a medium-FRET orientation (P = 0.51, for 224 ms) and jumped to a low-FRET orientation (P = 0.38, for 86 ms) before rotating to the high-FRET orientation (P = 0.74, for 273 ms).

### 3.2. FRET measurements of single $F_oF_1$-ATP synthases in solution using interleaved dual laser excitation

FRET analysis of conformational changes of single enzymes or motor proteins requires selective labeling of two marker positions in the protein. To facilitate selective labeling we used a genetic fusion construct of the autofluorescent protein EGFP to subunit *a* in the $F_o$ part[15]. Thereby every $F_oF_1$-ATP synthase is specifically labeled with the FRET donor that is excited at 488 nm. The second fluorophore is attached to a cysteine residue using a maleimide group at the fluorophore (Alexa-568-maleimide as the FRET acceptor). Accordingly we labeled the $F_o$ motor of ATP synthase with two fluorophores to monitor proton-driven rotation of the ring of *c* subunits[16], and we measured the rotation of the ε subunit with respect to the non-rotating *a* subunit[15].

Quantitative FRET distance information can be obtained from the ratio of the measured fluorescence intensities of FRET donor and acceptor in two spectrally distinct detection channels:

$$E_{FRET} = \frac{I_A}{I_A + \gamma I_D} \quad (2)$$

with $I_A$ and $I_D$, fluorescence intensities from the acceptor/donor dye, and γ, correction factor, which consists of the quantum yields of the dyes and the detection efficiencies of both detection channels.

Alternatively the lifetime information from the FRET donor dye can used, which requires excitation by a pulsed laser. An increasing FRET efficiency is reflected by a reduction of the donor lifetime

$$E_{FRET} = 1 - \frac{\tau_{DA}}{\tau_{D_0}} \qquad (3)$$

with $\tau_{DA}$, the fluorescence lifetime of the FRET donor in presence of the acceptor, $\tau_{Do}$ is the fluorescence lifetime of the donor in the absence of a FRET acceptor or other quenchers in the local environment. Using pulsed excitation allows to control the FRET efficiency calculation based on fluorescence intensities by the simultaneous determination of $E_{FRET}$ based on the single donor fluorophore lifetimes.

In our previous studies the EGFP fluorophore fused to $F_oF_1$-ATP synthase showed significant spectral shifts depending on the laser excitation power. Especially upon pulsed excitation, a second fluorophore population with red-shifted fluorescence and shorter fluorescence lifetime was found in EGFP-labeled $F_oF_1$-ATP synthases[15]. Intensity threshold criteria were applied that could reduce the fraction of the red-shifted fluorescent $F_oF_1$-ATP synthases in the histograms of FRET efficiencies. However, it was not possible to remove spectral shift-induced apparent FRET efficiency changes in the time trajectories. Accordingly these spectral shifts of the FRET donor appeared in the dwell time histograms of the FRET levels. Probing the existence of a FRET acceptor independently was the goal of the present approach.

An optimized pulsing scheme for the PicoTA 490 laser and the AOM for switching the 561 nm laser (for FRET acceptor probing) was developed. $F_oF_1$-ATP synthase labeled with EGFP at subunit *a* and with Alexa-568 at subunit ε were reconstituted into liposomes, and freely diffusing enzymes were detected in solution while traversing the overlayed laser foci for FRET donor and acceptor excitation (Fig. 6).

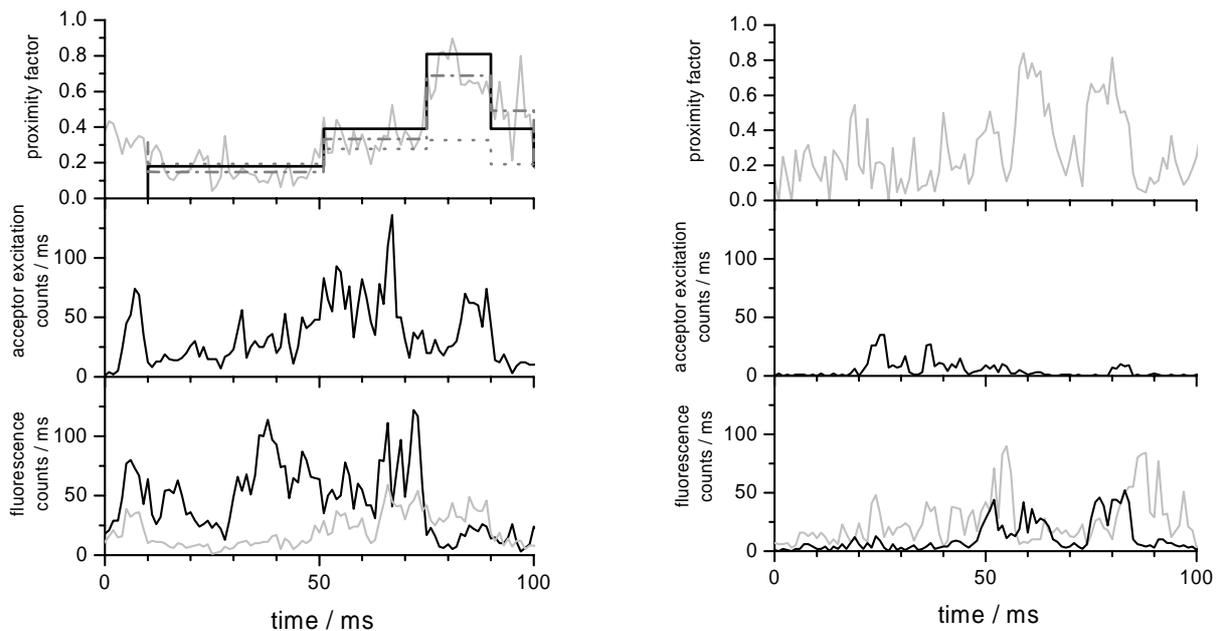

Fig. 6. Time trajectories of the fluorescence intensity changes of two $F_oF_1$-ATP synthases in the presence of 1 mM ATP Left, $F_oF_1$-ATP synthase showing FRET efficiency changes and rotation of the ε subunit. Right, $F_oF_1$-ATP synthase showing spectral fluctuation of the FRET donor fluorophore but not FRET efficiency changes as the FRET acceptor is lacking. The order of FRET efficiency changes as shown in the proximity factor trace above is low-FRET level → medium → high, as shown with the same FRET-labeled $F_oF_1$-ATP synthases buffer solution previously (but opposite to the FRET labeling for γ subunit rotation in Fig. 5). Upper trace left, the mean proximity factor for each FRET level (dash dotted) is compared to the lifetime-derived proximity factor (dotted line) for the HMM assigned states (solid line).

The pulse sequence synchronized the TCSPC electronics to have easy access to the FRET donor lifetimes by summing the fluorescence data created by the six consecutive pulses of the PicoTA laser (Fig. 3).

Software-based data analysis tools were applied for the FRET analysis. Single $F_oF_1$-ATP synthases were identified in the time trajectories using fluorescence intensity thresholds. Within the photon bursts FRET efficiency changes and the switching point of the FRET levels were determined by a Hidden Markov Model approach. In this work, we assigned a three state HMM with monoexponential dwell time distributions to the FRET data. One additional forth state was added for $F_oF_1$-ATP synthases without a FRET acceptor and to model those parts of the FRET time trajectories, where the acceptor dye is in a dark state. However, if the FRET donor is in a dark state (or missing), the fluorescence intensity is zero or at background level which makes a fifth state in the HMM model dispensable. In contrast to other Hidden Markov Models[17,18] applied to single-molecule measurements we replaced the Gaussians emission functions with Beta distributions[16].

Purified proteins for single-molecule fluorescence measurements contain a small amount of fluorescent impurities which yield short (due to binning up to 2 ms) photon spikes in the fluorescence intensity time trace. This will usually cause additional changes in the state sequence of the Markov Model. To prevent this, we added in front of each of the four single states of the Markov Model (for donor only, low FRET, medium FRET and high FRET) two extra states with a transition probability fixed to 1 and the same Beta distributions like the four main states. The transition matrix was implemented so that on each change in a new main state the Markov chain had to run over the two additional states. This yielded a minimum length of the dwell time of 3 ms without changing the parameters of the emission functions. The Beta functions were broadened by an exponent < 1 to match the broadening in the proximity factor distribution caused by time dependent changes in $\kappa^2$ and the photophysics of the dyes[19].

After the assignment of the HMM states for FRET level and dwell times, the lifetime of the FRET donor photons was calculated using a maximum likelihood estimator. For each FRET level the acceptor fluorescence intensity was checked by direct excitation. Relative intensities of FRET donor and acceptor were plotted against the FRET donor lifetime as shown in Fig. 7. As the consequence of the FRET acceptor probing in the time trajectories, the width of the lifetime distribution for each FRET level is narrowed after removal of the states without a fluorescent FRET acceptor. In addition, the red-shifted population of the EGFP in donor-only labeled ATP synthases, which was characterized by a lifetime of about 2 ns and a logarithmic intensity ratio of 0.6, was removed. The remaining FRET levels in rotating ATP synthases according to the HMM were found at a proximity factor P = 0.18 [$\log_{10}(I_D/I_A)$ = 0.664] for the low-FRET level, P = 0.39 [$\log_{10}(I_D/I_A)$ = 0.196] for the medium-FRET level, and P = 0.81 [$\log_{10}(I_D/I_A)$ = - 0.62] for the low-FRET level. The donor-only state was determined by P=0.098 [$\log_{10}(I_D/I_A)$ = 0.964].

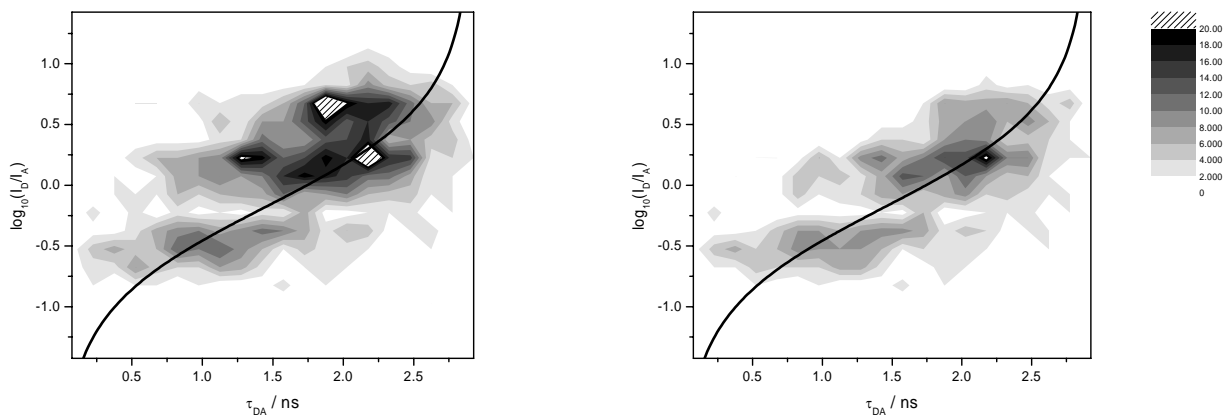

Fig. 7. Histogram of the logarithmic ratio of fluorescence intensities in the FRET donor and acceptor channel *versus* the FRET donor lifetime The black curve represents the FRET relation derived from equations (2) and (3). Left, all FRET levels of photon bursts showing 3 or more steps. Right, removal of the red-shifted population of donor-only states using the additional filter criterion of a mean fluorescence intensity from the acceptor probing channel higher than 6 counts per ms.

The dwell time histogram produced by all HMM assigned states was compared to the histogram containing only HMM states from photon bursts with 3 and more FRET levels, and to the histogram of rotating ATP synthases with an assured fluorescent FRET acceptor (as probed by direct excitation). In Fig. 8 the mean dwell times dropped from 48.8 ms for all HMM states to 17.6 ms for rotating $F_oF_1$-ATP synthases, and further to 15 ms for the rotating $F_oF_1$-ATP synthases with an existing FRET acceptor.

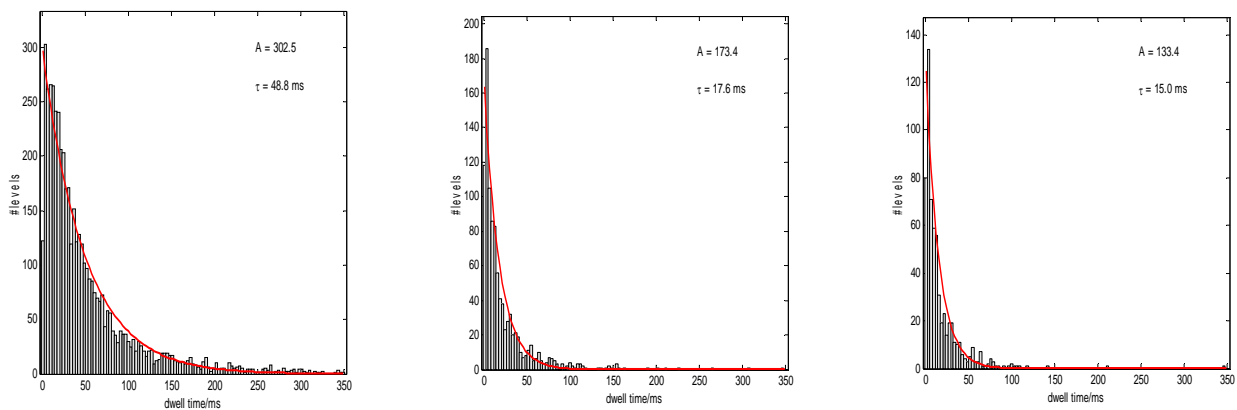

Fig. 8. dwell time histograms of the FRET levels created by the HMM. Left, all levels without filtering. Middle, all levels from bursts with 3 or more levels and without the first and the last level within a photon burst. Right, applied additional acceptor probing filter (at least 6 counts per ms on average) and without the first and the last level within a photon burst.

## 4. DISCUSSION

$F_oF_1$-ATP synthase exhibits rotary subunit movements during ATP hydrolysis and ATP synthesis. In order to gain detailed information about the mechanics and the mechanisms of the rotary motions, single enzymes have to be investigated. The single-molecule FRET approach to monitor subunit rotation by distance measurements within single membrane-embedded enzymes in real time had been refined here by two ways. Prolonged observation times using functionally immobilized liposomes containing $F_oF_1$-ATP synthase yielded trajectories of several seconds for a single enzyme. The FRET efficiency levels were found to be less noisy compared to solution data, where Brownian motion of the liposomes through the laser excitation focus caused strong fluorescence intensity fluctuations.

We expect to unravel the mechanism of ATP synthase inhibitors like aurovertin (see figure on the right), if a blocked rotation of subunits can be discriminated from a slow-down of the rotary motions. Biochemical rate determinations of ATP hydrolysis utilize a coupled enzyme reaction where the absorbance decrease of NADH is measured at 340 nm. The slope indicates the ATP turnover. The FRET labeled $F_oF_1$-ATP synthases, which were used for the surface immobilized FRET measurements, were found to hydrolyze more than 80 ATP molecules per second per enzyme. Addition of 20 µM aurovertin inhibited ATP hydrolysis by more than 80 percent. Preliminary single-molecule FRET data of the surface-immobilized $F_oF_1$-ATP synthases showed a significant amount of non-rotating enzymes.

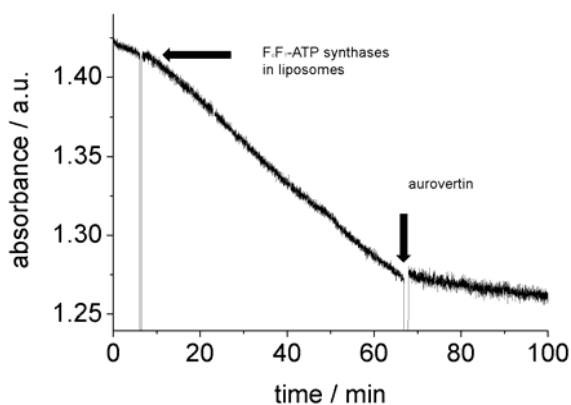

Fig. 9. ATP hydrolysis and aurovertin inhibition of 20 pM FRET-labeled $F_oF_1$-ATP synthases in liposomes. Addition of ATP synthase starts ATP hydrolysis, addition of aurovertin inhibits ATP hydrolysis.

Progress was made to eliminate the distortion by spectral fluctuations in single-molecule FRET data of freely diffusing enzymes. The FRET levels are affected by these spectral fluctuations resulting in wrong distances between the two fluorophore markers. Accordingly the assignment of conformational states is disturbed. The dwell times of the different conformational states of the $F_oF_1$-ATP synthase are also affected as spectral fluctuations are likely to occur on other time scales as the catalytic turnover.

In the current state of our software-based FRET data analysis, single $F_oF_1$-ATP synthases are identified in the FRET time trajectories by fluorescence intensity threshold criteria. Times of constant proximity factors in the FRET trajectory of the single molecule measurements were identified using Hidden Markov Models with emission functions based on modified Beta distributions. A four state model for the three conformational states and one photophysical state (FRET donor-only) was found to be sufficient to assign FRET levels and dwell times. Simultaneously present small fluorescent impurities could be tolerated by the model. The use of the FRET donor lifetime did not influence the HMM analysis. However, the FRET acceptor probing with the dual laser excitation scheme allowed for elimination of apparent FRET states caused by spectral fluctuations of the FRET donor. As the result, the remaining mean dwell times will now repesent the conformational dynamics of the $F_oF_1$-ATP synthases more accurately and are in good agreement with the biochemically determined turnover rates during ATP hydrolysis.

**Acknowledgements**

The authors want to thank Dr. B. Zimmermann and Prof. Dr. P. Gräber (University of Freiburg, Germany) for the gift of the FRET-labeled $F_oF_1$-ATP synthases in biotin-containing lipid vesicles which were used for the surface immobilized enzymes. We thank Y. Bi (University of Western Ontario, Canada) for support in constructing the plasmid for EGFP-fused $F_oF_1$-ATP synthases. We are grateful to *von Gegerfelt Photonics* and *Becker & Hickl* for the loan of the 561 nm laser Cobolt Jive and the synchronized TCSPC cards SPC-152. Financial support by the Deutsche Forschungsgemeinschaft (BO 1891/8-1, BO 1891/10-1) and the Landesstiftung Baden-Württemberg (network of competence 'functional nanodevices', http://www.nanonetz-bw.de) is acknowledged.